\documentclass[11pt]{article}
\usepackage{setspace}
\usepackage{amssymb}
\usepackage{amsmath}
\usepackage{amsfonts}
\usepackage{bm}

\def\be{\begin{equation}}

\def\ee{\end{equation}}

\def\dd{\partial}

\newcommand\of[1]{\left( #1 \right)}

\def\bea{\begin{eqnarray}}

\def\eea{\end{eqnarray}}

\newcommand\eps{\epsilon}

\begin{document}

\singlespace

\begin{flushright} BRX TH-627 \\
CALT 68-2814
\end{flushright}

\vspace*{.3in}

\begin{center}

{\Large\bf No local Maxwell duality invariance}

{\large S.\ Deser}

{\it Physics Department,  Brandeis University, Waltham, MA 02454 and \\
Lauritsen Laboratory, California Institute of Technology, Pasadena, CA 91125 \\
{\tt deser@brandeis.edu}
}

\end{center}

\begin{abstract}
I show, in canonical formulation of flat space Maxwell theory, that its duality rotation invariance must have constant parameter, cannot be promoted to a local one by adding a compensating field, and that these conclusions hold in the curved space Maxwell-Einstein extension.
\end{abstract}

\section{Introduction}

While invariance of source-free Maxwell theory under electric-magnetic rotation is ancient lore--
it is perhaps the oldest field-theoretical duality--a proper demonstration of its off-shell validity in terms of the theory's degrees of freedom, rather than just the formal interchangeability of the field equations $\dd_\mu \, F^{\mu\nu}=0=\dd_\mu\ ^*F^{\mu\nu}$, is relatively recent [1,2]. [Off-shell duality of the action is hardly manifest, since it is a difference, $I=\int d^4x\,  [{\bf E}^2-{\bf B}^2]$ rather than a sum, 
of squares, nor is it, also contrary to intuition, invariant under hyperbolic rotation.] The (admittedly unphysical) possibility of promoting this constant rotation parameter to a space-time function, was recently [3] raised and excluded, using a Noether procedure in the theory's two-potential formulation. It motivates  the present derivation of this no-go result within the original canonical formulation of [2], as well as showing that a compensating field cannot avoid it. I will then extend these conclusions to the generally covariant  theory, noting that they agree with the ``already unified", purely metric, formulation of Einstein-Maxwell [6],  which is also only constant duality invariant. 

\section{Flat space}

We begin by noting two formal arguments against a variable rotation angle $\alpha$: First, the two $\dd F=0= \dd\ ^*F$ equations would obviously not transform into each other under rotation of $F_{\mu\nu}$ and its dual field strength $\ ^*F^{\mu\nu} \equiv 1/2\,  \eps^{\mu\nu\alpha\beta} \, F_{\alpha \beta}$, but leave residues $\sim( \dd \, \alpha)\, (F, \ ^*F)$; similarly, the Lagrangian $F^2$ would rotate into $\alpha \, F\ ^*F$, which is only a total divergence for constant $\alpha$. These arguments are purely formal because for any putative transformation to be meaningful--let alone an invariance--it must be explicitly implementable in terms of the theory's true degrees of freedom. At that basic level, duality is simply a canonical transformation, rotating the two conjugate helicity degrees of freedom according to $(p^a, q^a)\longrightarrow \eps^{ab} \, (q_b,-p_b)$, which is why duality invariance also applies to all free spin $>1$ gauge fields [4].
We first consider the free flat-space Maxwell action after solving its Gauss constraint $\dd_i\, E^i=0$, when the longitudinal pure gradient component of ${\bf E}$ like that of ${\bf B}$ vanishes; the system then depends only on the gauge-invariant independent conjugate transverse, pure curl, pair $({\bf E}^{T}, {\bf A}^{T})$; we drop the ``$T$" superscript henceforth,
\begin{equation}
   I [{\bf E},{\bf A}]= 1/2\,  \int d^4x \, [(-{\bf E} \cdot \dot {\bf A} -1/2\, ({\bf E}^2+ {\bf B}^2)], \, \, \, \, \, \, \,   B^i({\bf A}) \equiv \eps^{ijk} \dd_j\,  A_k,           
   \end{equation}
   since the contraction of any transverse--longitudinal 3-vector pair vanishes upon spatial integration.  This reduced action (1) is off-shell invariant under the infinitesimal canonical transformation with constant infinitesimal parameter $\alpha$,
\begin{equation}
\delta {\bf E}= \alpha \, {\bf B}({\bf A}), \, \, \, \, \, \, \, \delta {\bf A}=\alpha \nabla \times  {\bf E}/ (\nabla^2) \leftrightarrow \delta {\bf B}({\bf A})=-\alpha \, \bf{E},     
\end{equation}
generated by  
\begin{equation}
G[{\bf E}, {\bf A} ]=1/2\,  \int d^3x \, [{\bf E} \, \nabla^{-2}\, \cdot \of{ \nabla \times  {\bf E}} - {\bf A} \cdot  \of{\nabla \times {\bf A}}].                        
\end{equation}
Rotation invariance is manifest for the Hamiltonian's sum of squares, while the kinetic term varies into $\alpha$ times the sum of time derivatives of two ($D=3$) Chern-Simons-like terms, $\delta {\bf E}\cdot \dot {\bf A} \sim \alpha \, d/dt \of{ \nabla \times {\bf A} \cdot {\bf A} }$, hence vanishes by parts integration, and similarly for ${\bf E} \cdot \delta \dot {\bf A}$. Note, 
for the latter, that the non-local Coulomb propagator in (2) doesn't spoil integrations by parts.
If we try to promote the scalar $\alpha$ to a function $\alpha(x)$, invariance of (1) under (2) is lost, the result being essentially 
\begin{equation}
\delta I= \int d^4x \, \alpha(x)\,  F\ ^*F= -\int d^4x \, \dd_\mu \, \alpha(x)\, C^\mu,              
\end{equation}
using the identities 
$F\ ^*F \equiv \dd_\mu C^{\mu} \equiv \dd_\mu \, \eps^{\mu \nu \alpha \beta}\,   A_\nu\, F_{\alpha\beta}$.          
The $C^\mu$ are the Chern-Simons densities of the $3$-spaces orthogonal to $\mu$, expressed in terms of $({\bf E},{\bf A})$. This quantifies the original naive no-go argument.  A more direct and telling one is the following.  As stated above, the Maxwell action depends only on transverse modes, so there is an unavoidable contradiction between any local version of (2) and transversality: Even if one keeps $\alpha(x)$ inside the various curls, both in $\delta {\bf E} \rightarrow \nabla \times (\alpha\, {\bf A})$, and in $\delta {\bf A}$, this introduces derivatives of the rotation angle, destroying the direct ${\bf E} \leftrightarrow {\bf B}$ and with it the whole rotation group basis.

A final attempt at localizing is to introduce a compensating vector field $W_\mu$ that gauge-transforms by $\dd_\mu \alpha(x)$, and couple it to $C^\mu$. While this variation indeed compensates that of (2), as designed, there instead appear new $W_\mu$-dependent terms from the variation of $C^\mu$. For example, $C^0 \sim {\bf A} \cdot {\bf B}+\ldots$ , whose variation is $\sim -2\, \alpha\, {\bf A} \cdot {\bf E} +\ldots$, and likewise for the ${\bf C} \sim \alpha\, {\bf A} \times {\bf E}$. 
No-go theorems have been historically interesting in physics partly
because they encourage
searching for loopholes. Here, after emphasizing the no-go without
compensating fields, also
in the gravitational extension below, we went on to exclude the most
obvious compensator
candidate, a gauge vector potential. At the very least, ruling out
that standard approach to
localizing invariances considerably raises the ``loophole bar".   This completes the canonical no-go result for flat space Maxwell theory.

\section{Einstein-Maxwell}

As shown in [1], the covariant Maxwell action also remains constant duality invariant. Briefly,    
the covariant extension of the Lagrangian (1), in  ADM $3+1$ metric  notation, [5] 
\begin{equation}
 L= -E^i \, \dot A_i - 1/2 \, N\,  g^{-1/2} g_{ij} (E^i \, E^j+ B^i \, B^j) + \eps_{ijk} \, N^i \, E^j \, B^k.    
\end{equation}
is invariant after using the (unchanged) Gauss constraint; it  again implies that ${\bf E}$ is transverse; instead of using the ``$T$" notation, it is more useful here to express ${\bf E}$ in terms of an auxiliary 
$3$-vector ${\bf C}$,
\begin{equation}
E^i \equiv \eps^{ijk}\,  \dd_j \, C_k.  
\end{equation}
This definition makes the ${\bf E}$-${\bf B}$ symmetry manifest; of course $B^i=\eps^{ijk} \dd_j \, A_k$ still. Duality is now simply a rotation in $({\bf A},{\bf C})$ space, one that obviously maintains the ${\bf E} \leftrightarrow \bf{B}$ rotation (2). This means in turn that not only the energy density $({\bf E}^2 + {\bf B}^2)$, but all stress tensor components, 
$(E^i\,  E^j + B^i\,  B^j)$, the momentum density ${\bf E}\times{\bf B}$, which rotates into ${\bf B}\times {\bf B} -{\bf E}\times{\bf E} \equiv 0$, as well as the kinetic term ${\bf E} \cdot \dot{\bf A}$, are invariant. The latter now takes the form
\begin{equation}
{\bf E} \cdot \dot {\bf A} = \eps^{ijk}\,  \dd_j \, C_k\,  \dot A_i 
\end{equation}
and varies into  
\begin{equation}
\alpha \, \eps^{ijk} \, [-\dd_j \, A_k\,  \dot A_i + \dd_j \, C_k\,  \dot  C_i ]= \alpha \, \dd_0\,  [CS ({\bf A})-CS({\bf C})],
\end{equation}
a difference of Chern-Simons densities. For constant $\alpha$ (only), the integral vanishes, insuring invariance. A compensating field $W_\mu$ cannot change this, just as in flat space: Its $W_0$ component would have to couple to the CS difference in (8), and the latter's variation is non-zero,
being essentially a ``cross-CS" term $\sim \eps^{ijk}\,  A_i \, \dd_j \, C_k$. [We emphasize that a constant scalar parameter, unlike vectorial or spinorial ones, is perfectly meaningful in curved space.]

Finally, we comment on the connection between covariant duality invariance and the coupled Maxwell-Einstein system. As we saw, all components of the Maxwell tensor,
\begin{equation}
T_{\mu\nu} = F_{\mu \alpha}\,  F_{\nu}^{\, \, \, \alpha} +\ ^*F_{\mu\alpha} \, \ ^*F_\nu^{\, \, \, \alpha} = T_{\nu\mu}, \, \, \, \, \, \, \,  T_\mu^\mu \equiv 0,  
\end{equation}
are duality invariant; indeed, being algebraic in the field strength, they are even (if only formally) locally invariant, a fact relevant below. However, Maxwell's equations are as essential a part of the coupled system as Einstein's, being required to ensure that $T_{\mu\nu}$ is (covariantly) conserved, and hence is an allowed source of the Einstein tensor. The Maxwell tensor's two special algebraic properties-- tracelessness and idempotence--can be imprinted on the Einstein tensor, whereupon the Einstein equations can be restated in purely metric terms, but only up to a local ambiguity between ${\bf E}$ and ${\bf B}$ (or $F$ and $\ ^*F$), as expected since these equations only see the locally invariant (9). This is the content of the famous ``already unified" [6] form of Einstein's equations with a Maxwell source: the Maxwell aspect is--almost, but not quite--subsumed in purely metric terms. This is commonly thought to mean that the theory is locally rotation invariant. However, it only follows that the metric variables can characterize its sources at any point just up to a parity ambiguity; Maxwell's equations, essential to the validity of the source (9), still remain in the game and they only enjoy constant duality invariance, all in accord with our underlying canonical considerations.

\section{Conclusion}

Local Maxwell duality invariance is excluded both in flat and curved space and cannot be implemented by introducing a compensating field.  This constant invariance agrees with that of the Maxwell-Einstein system's ``already unified" form.

I thank M Henneaux for correspondence. This work was supported by NSF PHY 07-57190 and DOE DE-FG02-164 92ER40701 grants.

NOTE ADDED. My submission has prompted a subsequent posting [7] explicitly excluding an earlier [8], purportedly successful, localization involving a
cascade of compensating vector and scalar auxiliaries.

\end{document}